\documentclass[aps, prd, amsmath, floats,floatfix,twocolumn, superscriptaddress]{revtex4-2}
\usepackage{hyperref}

\newcommand{\hh}{{\hspace{.3mm}}}

\def\sideremark#1{\ifvmode\leavevmode\fi\vadjust{\vbox to0pt{\vss
 \hbox to 0pt{\hskip\hsize\hskip1em

 \vbox{\hsize3cm\tiny\raggedright\pretolerance10000
 \noindent #1\hfill}\hss}\vbox to8pt{\vfil}\vss}}}%
                        
\usepackage{mathrsfs}
\usepackage{natbib}
\usepackage{amsmath}
\usepackage{amssymb}
\usepackage{lmodern}

\usepackage[T1]{fontenc}
\usepackage[latin9]{inputenc}
\usepackage{geometry}
\usepackage{subfigure,lmodern, amsmath,amssymb, graphicx, pifont, adjustbox, bm, xcolor}
\usepackage{amsfonts}
\usepackage{geometry}
\usepackage{comment}
\usepackage{mathtools}
\usepackage{float}
\usepackage{slashed}
\usepackage{ragged2e}

\usepackage{bm}

\newcommand{\rchi}{\mu}

\begin{document}

\title{Stress and Geometry for Isotropic Singularities}

\author{ A. Rod Gover}

\affiliation{
  Department of Mathematics,
  The University of Auckland,
  Private Bag 92019,
  Auckland 1142,
  New Zealand}
  \email{r.gover@auckland.ac.nz}

\author{Jaros\l aw Kopi\'nski}
 
  \affiliation{
  Center for Theoretical Physics, Polish Academy of Sciences, \\ 
  Al.\!\! Lotnik\'ow 32/46, 02-668 Warsaw, Poland} \email{jkop@math.ucdavis.edu}
  \affiliation{
  Center for Quantum Mathematics and Physics (QMAP),
  Department of Mathematics,
  University of California,
  Davis, CA95616, USA} 
  \email{wally@math.ucdavis.edu}

\author{Andrew Waldron}
  \affiliation{
  Center for Quantum Mathematics and Physics (QMAP),
  Department of Mathematics,
  University of California,
  Davis, CA95616, USA} 



\date{\today}

\begin{abstract}

We develop the mathematics needed to treat the interaction  
of geometry and stress at any isotropic spacetime singularity. This enables us to  handle the Einstein equations at the initial singularity and characterize 
allowed
 general relativistic stress-energy tensors. Their leading behaviors  are dictated by an  initial hypersurface  conformal embedding. We  also show that an isotropic  Big Bang determines a canonical non-singular metric on and about the initial hypersurface as well as a  cosmological time.  This assigns a volume and energy  to the initial  point singularity.

\end{abstract}


\maketitle

\section{Introduction}
Is the causal structure of our universe singular at the Big Bang? This question is of physical import since, as we shall show, a  well-defined causal structure at the initial Big Bang singularity imposes strong constraints on the matter content of the early universe.

We perform our analysis in the context of spacetimes with isotropic singularities. As explained in~\cite{goodewan, tod1}, an  isotropic singularity can be removed by multiplying the physical metric by some power of a suitable timelike coordinate; see Section~\ref{IS}. The conformal structure can then be extended across the initial singularity, which is now described by a spacelike hypersurface.
This means that all physical spacetime directions contract at the same rate when approaching the singularity. 
This definition need not imply any particular isometries for some choice of spatial slices.
It includes the 
standard Friedmann--Lema\^itre--Robertson--Walker (FLRW) example discussed below.
Isotropic singularities are relevant in light of
Penrose's Weyl Curvature hypothesis~\cite{penrose} asserting that the Weyl tensor is finite  at any initial singularity even if the Ricci curvature is singular~\cite{f1}. They have also been studied when subject to various  underlying matter model assumptions~\cite{goodewan, tod1, tod2, newman, tod3, tod4, ang, tod5, tod6}. Our analysis applies to generic stress-energy tensors (stress) for spacetimes with isotropic singularities.

Because the metric along an isotropic singularity is degenerate but the conformal structure is still well-defined, early universe physics is dictated by the mathematics of conformally embedded hypersurfaces. Conformal submanifold embeddings are crucial to
 the theory of observables in the AdS/CFT correspondence~\cite{GrahamWitten,Ryu}.
This machinery can be fruitfully applied to cosmology.
Indeed, any study of causal structures 
amounts to a problem in 
 confornal geometry. There is a well established ``tractor calculus''  for  handling conformal geometries on which we rely.
 Our presentation is
self-contained, though many key definitions 
are relegated to
 footnotes; excellent resources for further details include~\cite{BEG,Curry}.

\section{Conformal Geometry}

\newcommand{\bg}{{\bm g}}

For simplicity we  focus on generic, dimension~4~\cite{f2}, causal structures given by the data of a Lorentzian conformal geometry $(M,\bg)$~\cite{Senovilla}, where~$\bg$ denotes a conformal class of metrics~$g$ with equivalence given by  rescalings $ \Omega^2g\sim g$ for $0<\Omega\in C^\infty M$. Parallelism determined by the Levi--Civita connection $\nabla$ is a central mathematical construct of general relativity. Its  conformal geometry generalization, known as the tractor connection~$\bm \nabla$, promotes the tangent bundle~$TM$  to a ``tractor bundle''~${\bm T}M$ with dimension six  fibers~\cite{Thomas,BEG}. 
Tractors are basic objects for theories incorporating   local conformal transformations and diffeomorphisms~\cite{Curry}. 
They are even useful for analyzing 
systems that are not invariant under local Weyl symmetry.
Indeed~$\bg$ contains  
 a metric solving the ($\Lambda$-)vacuum Einstein equations precisely when there is a parallel tractor vector field $I\in {\bm T}M$~\cite{BEG, GovNu} (details are discussed just before Eq.~\eqref{freeyourstress} below), {\it viz}
\begin{equation}
\label{parallel}
{\bm \nabla}I=0\, .
\end{equation}
To incorporate Lorentz symmetry in physical theories, 3-vectors are promoted to 4-vectors. Tractors promote 4-vectors to 6-vectors to manifest conformal symmetry.
Given a choice of metric $g\in \bg$, a 
 tractor~$I$ is a  triple
\begin{equation}\label{transform}
I:\stackrel g=
\scalebox{.9}{$\begin{pmatrix}\sigma\\n^b\\ \rho\end{pmatrix}\stackrel{\Omega^2 g}=\begin{pmatrix}\Omega \sigma\\ \Omega(n+ \sigma d\log \Omega)^b\\\Omega^{-1}(\rho -{\mathcal L}_n\log\Omega-\tfrac\sigma2 |d\log\Omega|_g^2)
\end{pmatrix}$}\, ,
\end{equation}
where 
$\sigma$, $\rho$ are scalars, ${\mathcal L}_n$ is the Lie derivative along the vector $n$, and the  gauge transformation in Eq.~\eqref{transform} is valued in the parabolic subgroup (preserving a lightlike ray) of the spacetime conformal group $SO(4,2)$.
Tractors of any tensor rank are also well defined~\cite{BEG,Curry}.
The tractor connection is defined by
\begin{equation}\label{Istaythesame}
{\bm \nabla}_a  I:=
\scalebox{.9}{$
\begin{pmatrix}\nabla_a{\bm \sigma} - n_a \\
 \nabla_a n^b + \bm{\sigma} P_a{}^b+\rho\hh \delta_a{}^b\\\nabla_a \rho - n_c P_a{}^c\end{pmatrix}$}\, .
 \end{equation}
In the above $\bm \sigma$ denotes a conformal density of weight~$1$.
A weight $w$ conformal density~\cite{f3} 
may be viewed as a power of a volume form, so
is a section of $ \big[(\wedge^4 TM)^2]^{\frac w{8}}=:{\mathcal E}M[w]$. This is a power of a tensor density, so  the Levi-Civita connection is well-defined acting upon it. A density $\bm \sigma$ may also be understood as an equivalence class of metric-function pairs $(g,\sigma)\sim (\Omega^2 g,\Omega\sigma)$. Indices are raised and lowered using $\bg$ and $\nabla$ is the Levi-Civita connection (see~\cite{Curry}). Also~$P$ denotes the Schouten tensor and~$J$ its trace.

The standard tractor bundle ${\bm T}M$ comes equipped with a parallel ``tractor metric'' $h$ and (unlike the tangent bundle) a canonical tractor vector field $X\in {\bm T}M[1]$~\cite{BEG, Go,Curry}.
Indeed,  $h(I,X)=\bm \sigma\in {\mathcal E}M[1]$ for any $I$ as given  on the left hand side of Eq.~\eqref{transform}; this defines $X$.
Moreover, when Eq.~\eqref{parallel} holds, it turns out that the vacuum cosmological constant $\Lambda_{\rm vac}$ obeys 
$$- \tfrac13 \Lambda_{\rm vac}
= 2 \bm \sigma \rho + |n|_{\bm g}^2
=h(I,I) =:\hh I \!\hh\cdot \!\hh I\hh=:I^2\, .
$$

To incorporate matter, we must couple  stress  to the right hand side of Eq.~\eqref{parallel}, since it is a conformally covariant reformulation of Einstein's equations in vacua.
On the other hand, when the function $\sigma$ is a good coordinate for some hypersurface $\Sigma$,  the local conformal embedding data $\Sigma\hookrightarrow (M,\bg)$ is encoded  by ${\bm \nabla} I\in TM\otimes {\bm T}M$. The tractor ${\bm \nabla} I$ is a canonical conformal extension of the extrinsic curvature.
It follows that there is a natural correspondence between  stress and local conformal embedding data.

\section{Isotropic Singularities}\label{IS}
An isotropic singularity is a spacelike hypersurface $\Sigma$
in a spacetime $M$ with a degenerate 
 physical metric $\check g$ such that, for $\alpha<0$ and any defining function $\tau$~\cite{f4}, 

\begin{equation}\label{checksarehere}
g=\tau^{2\alpha} \check g 
\end{equation}
extends  to a smooth metric across $\Sigma$. The degree of metric singularity for the (zero $\Lambda$)
conformally flat FLRW
 spacetime with
perfect fluid pressure to density ratio $\kappa$
is $$\alpha_{\rm FLRW}=-\tfrac{2}{3\kappa +1}\, .$$
Even this simplest of cosmological scenarios 
allows non-integer $\alpha$.
This parameter controls
 both smoothness of the physical metric~$\check g$ and the volume expansion rate.

Any other {\it bona fide} metric $g'=\Omega^2 g$ corresponds to  a rescaled defining function $\tau'=\Omega^{\frac1\alpha}\tau$. 
Thus we may write~\eqref{checksarehere} as
 $$
\check g = {\bm \tau}^{-2\alpha} \bg\, ,
$$
with ${\bm \tau}\in {\mathcal E}M[\frac1\alpha]$, where
no
 particular choice of metric in the conformal class~$\bg$ has been made.
Indeed the causal structure of $\bg$  is well-defined across the initial singularity.

\medskip

Our aim is to analyze the Einstein field equations
\begin{equation}\label{themostimportantequation}
\check G +\Lambda \check g= \check T\, ,
\end{equation}
where $\check T$ is the stress of a universe with cosmological constant $\Lambda$ and degenerate physical metric~$\check g$. 
For this we use maps from weight 1 scalar densities to weight~0 tractors and from 
tractor-valued one-forms to weight~1
 symmetric trace-free tensors~\cite{Eastwood,Curry} (denoted by \raisebox{-1.3mm}{$\, \mathring{}\, $})
\begin{equation}\label{sctr}
{\bm \rchi}\!\stackrel I\longmapsto\!
\!
\scalebox{.8}{$\begin{pmatrix}
\bm \rchi\\
 \nabla^b \bm\rchi \\\!\:
-\tfrac14(\square \!+ \!J){ \bm \rchi}\!\hh
\end{pmatrix}$}
\, ,\qquad\:
\scalebox{.8}{$
 \begin{pmatrix}
0\\
 \mathring x_a{}^b \\\!
-\tfrac13 \nabla_b \mathring x_a{}^b
\end{pmatrix}$}
 \stackrel {q^*} 
 \longmapsto \!
 \mathring x_{ab}\, .
\end{equation}
When $\bm \rchi$ is non-vanishing
almost everywhere, $I_{\bm \rchi}$ is termed a {\it scale tractor}. 
For {\it any} weight one density~$\bm \rchi$, 
the definitions of~$\bm \nabla$ and~$I_{\bm \rchi}$ in Eq.s~\eqref{Istaythesame} and~\eqref{sctr} imply vanishing of the top slot of~$\bm\nabla I_{\bm \rchi}$. So by virtue of Eq.~\eqref{transform} its middle slot is a trace-free conformally covariant rank two tensor  equaling~$q^* \bm\nabla I_{\bm \rchi}$.
Remembering that the $\mathring G=2\mathring P$, it follows from Eq.~\eqref{Istaythesame} that~$2\bm \rchi^{-1} q^*  \bm \nabla I_{\bm \rchi}$ is precisely the trace-free Einstein
tensor for the metric $\bm \rchi^{-2} \bg$, wherever this is defined. This explains the relationship between parallel scale tractors and Einstein metrics.
Eq.~\eqref{themostimportantequation} now reads 
\begin{eqnarray}\label{freeyourstress}
q^*{\bm \nabla} 
I_{\bm \tau^\alpha}
&=&\, \,\;
\tfrac{{\bm \tau}^\alpha}{2} \hh\mathring{\check T}\, ,\\[1mm]
\label{traceyourstress}
   I^2_{\bm \tau^\alpha} \, \, \, &=&\tfrac1{12}\check T_a{}^a- \tfrac13  \Lambda
\, .
\end{eqnarray}
Scale tractors  are potentials for Einstein's equations since the derivative of $I$ yields trace-free stress. Reference~\cite{GoWeyl} shows that  the Einstein--Hilbert action is the integral of $I^2$.
Note that it already follows from Eq.~\eqref{traceyourstress}
that the trace of the stress  for a spacetime with isotropic singularity cannot vanish along~$\Sigma$ unless $\alpha\!=\!-1$,  as the leading behavior of $I^2_{\bm \tau^\alpha}$ is 
$\frac{\alpha(\alpha+1)}2\bm \tau^{2\alpha-2}|\nabla\bm \tau|^2_{\bg}$.

\section{Traversing the singularity}

Spacetimes whose singularities are isotropic admit a global causal structure.
Hence, even though the  Einstein tensor is singular across an isotropic singularity, there are a number of well-defined geometric quantities, invariant to the structure (meaning that they are determined by the structure alone), that constrain matter.
The Weyl tensor~$W_{ab}{}^c{}_d$ is defined independently of any choice of~$g\in \bg$ but, unlike the conformally covariant Bach tensor $B_{ab}:=\square P_{ab} - \nabla^c \nabla_a P_{bc} + P^{cd}W_{acbd}
$, it is not related to stress by a local differential operator. Let~${\mathscr P}$ be
 the conformally covariant {\it partially massless wave-operator } defined, acting on a weight~$1$ trace-free symmetric tensor $\mathring x_{ab}$,
by~\cite{PM}
$$
{\mathscr P}\hh \mathring{x}_{ab} :=  \square\hh  \mathring{x}_{ab} \!-\! \nabla_c \nabla_{(a} \mathring{x}^c{}_{b)_\circ} \!-\! \tfrac{1}{3} \nabla_{(a} \nabla_{|c|} \mathring{x}^c{}_{b)_\circ} \!+\! W_{a}{}^c{}_b{}^d \mathring{x}_{cd} \, .
$$
For any non-vanishing weight one density $\bm \rchi$~\cite{GSS},
\begin{equation}\label{bach}
\bm \rchi\hh  B:={\mathscr P}q^*
\bm \nabla I_{\bm \rchi}\, .
\end{equation}

Given a causal structure $\bg$, the Bach tensor is non-singular so the above implies that the physical stress obeys a d'Alembert-type equation~\cite{KoVa}  
 \begin{equation}\label{B2stress}
 B=\tfrac1{\bm \tau^{\alpha}}\hh {\mathscr P}\hh  \big( 
 \tfrac{\bm \tau^\alpha}2 \mathring {\check T}\big)\, .
 \end{equation}
 The Bach tensor is a natural invariant of a conformal structure~\cite{Bach}, so the above relates causality and stress.

\subsection{Singularity Geometry}

So far  the
conformal embedding data
$\Sigma\hookrightarrow (M,\bg)$ determined by the isotropic singularity
 has not been used. 
This data 
determines uniquely the local asymptotics of another metric~$g_+$, termed the {\it singular Yamabe metric}, whose scalar curvature obeys 
\begin{equation}\label{theYamsings}
R^{g_+}=12 + {\mathcal O}(\sigma^4)\, ,
\end{equation}
where $\sigma$ is any defining function for $\Sigma$. 
Interestingly enough, the initial hypersurface $\Sigma$ is a conformal infinity of 
 $g_+$.
An all order  ``singular Yamabe problem''~\cite{Loewner,Aviles,Mazzeo,ACF} solution amounts to finding~$\bm \sigma=[g,\sigma]\in {\mathcal E}M[1]$ such that $I_{\bm \sigma}^2=-1$
for which $g_+:=\bm \sigma^{-2} \bg$.
The expansion coefficient 
of the~$\bm\sigma^4$ term  in $I_{\bm \sigma}^2+1$, along $\Sigma$,
is a weight~$-4$ conformal hypersurface invariant~\cite{ACF,will1,will2} equaling  the variation of an energy functional $E_\Sigma$~\cite{gra,renvol,Glaros}. This   energy  is the anomaly in the renormalized volume of $(M,g_+)$~\cite{renvol,Glaros}  and a conformal invariant of the initial singularity
$$
E_\Sigma = \int_\Sigma \mathring K_{ab} \mathring F^{ab}\, 
 dV_{g_\Sigma}
\, . 
$$ 
The above integral is over any metric in the conformal class of metrics $\bg_\Sigma$ induced along $\Sigma$ by~$\bg$. It is invariantly defined because the contraction of the trace-free extrinsic curvature $\mathring K$ with the {\it Fialkow tensor}~\cite{Fialkow,Stafford},
$$
F_{ab}:=  \hat n^c \hat n^d W_{cabd} - \mathring{K}_{ac} \mathring{K}_{b}{}^c +\tfrac 14 \mathring K_{cd} \mathring{K}^{cd} \overline{g}_{ab} \in \odot^2T^*\Sigma\, ,
$$
defines a conformal density of weight $-3$. The 
extrinsic curvatures $(K,F)$ have respective  transverse orders~$(1,2)$  and are  termed 
second and third fundamental forms~\cite{f5}.  
They give
 the first two elements in a sequence of trace-free conformal hypersurface invariants defined along $\Sigma$ and termed {\it conformal fundamental forms}~\cite{CFF}. 
These are conformally invariant obstructions to the problem of finding an asymptotically de Sitter  (dS) metric with conformal infinity $\Sigma$. They probe derivatives of~$\bg$ off of~$\Sigma$ in the direction of the (future-pointing) timelike unit normal $\hat n\in TM[-1]|_\Sigma$.

The extrinsic curvature $K$ measures the difference between the Levi-Civita connections $\nabla$ and $\bar \nabla$ of $M$ and~$\Sigma$ respectively, while the Fialkow tensor measures that of the 
respective tractor connections $\bm \nabla$ and $\bar {\bm \nabla}$ of $\bg$ and~$\bg_\Sigma$~\cite{Stafford}.

The    {\it fourth conformal fundamental form}~\cite{CFF,GoKo} takes three normal derivatives of~$\bg$~\cite{f6}, \begin{multline*}
\mathring L_{ab}:= \left( \hat{n}^c C_{ c (ab)} \right)^\top + H \hat n^c \hat n^d W_{acbd}
\\
- \bar \nabla^c (\hat n^d W_{d(ab)c})^{\!\top}
\!\!
\in\! \circledcirc^2 T^*\Sigma[-1]\, .
\end{multline*}
Here $C_{abc}:=2\nabla_{[a} P_{b]c}$ is the Cotton tensor and 
$\circledcirc$ denotes the trace-free symmetric product of one forms. The tensor~$\mathring L$ is distinguished in the context of dS$_4$ metrics, because rather than obstructing solutions, it extracts the second piece of boundary data (the first being~$\bg_\Sigma$).

The locally determined singular Yamabe asymptotics of~\eqref{theYamsings} terminate at order four, so the fourth conformal fundamental form $\mathring L$ is the last tensor determined this way. 
Before using the geometric triple $(\mathring K,\mathring F,\mathring L)$ to constrain \scalebox{.85}{$\mathring{\check T}$}, we study further (pseudo-)Rie\-mannian data determined by the isotropic singularity.

\subsection{The Big Bang Metric}

Remarkably there is  a canonical 
Riemannian metric along  the big bang hypersurface $\Sigma$.
It is  constructed from the solution $\bm \sigma$ to the singular Yamabe problem:
The weight~1 density $\bm \gamma$ defined by
\begin{equation}\label{ratio}
\bm \gamma^{\frac1\alpha -1}:= \frac{\bm\tau}{\bm\sigma}
\end{equation}
is nowhere vanishing by our smoothness assumptions and therefore defines a Lorentzian metric
\begin{eqnarray*}
g_{\bm \gamma} := \bm \gamma^{-2} \bg\, ,\\[-2mm]
\end{eqnarray*}
\noindent
and in particular a Riemannian metric $g_\Sigma$ on the spacelike isotropic singularity $\Sigma$~\cite{f7}.
Thus the Riemannian three manifold $(\Sigma, g_\Sigma)$ is an invariant of the   Big Bang, as is its volume $
V_\Sigma=
\int_\Sigma dV_{g_\Sigma}
$.

\smallskip

Because
the  ``Big Bang metric'' $g_{\bm \gamma}$ is an everywhere smooth  element of the conformal class $\bg$,  it determines the triple of conformal fundamental forms $(\mathring K, \mathring F, \mathring L)$ by the formul\ae\  
above.
Importantly the metric $g_{\bm \gamma}$ also furnishes the early universe with a canonical cosmological time coordinate
$$
t:=\frac{\bm \sigma}{\bm \gamma}
$$
depending on data of the isotropic singularity alone. It is very useful for analyzing physical stress at a Big Bang singularity.

\section{Big Bang Stress}

The behavior of stress at an isotropic singularity can be studied using potentials $I_{\bm\tau^\alpha}$, $I_{\bm \sigma}$ and $I_{\bm \gamma}$.
By virtue of Eq.~\eqref{ratio} and the definition of a scale tractor in Eq.~\eqref{sctr}, these must be related:
\begin{widetext}
\begin{equation}\label{formula}
 I_{\bm \tau^{\alpha}}=
t^{\alpha-2}\Big[-
\tfrac{\alpha(\alpha-1)}{4\bm \gamma}
I_{\bm \sigma}^2
X
+t\big(
\alpha I _{\bm \sigma}+\tfrac{\alpha(\alpha-1)}{2\bm \gamma} I_{\bm\sigma}.I_{\bm \gamma}\hh X
\big)
+t^2\big(
(1-\alpha)I_{\bm \gamma}
-\tfrac{\alpha(\alpha-1)}{4\bm \gamma}I_{\bm \gamma}^2 \hh X
\big)\Big]\, .
\end{equation}
\end{widetext}
This is the fundamental equation governing 
stress at an isotropic singularity, all quantities above are determined by the basic geometric data $\check g$ or equivalently $(\bg,\bm \tau)$.

\subsection{Trace of Stress}

The square of a scale tractor measures scalar curvature/trace of stress (see Eq.~\eqref{traceyourstress}) so  Eq.~\eqref{formula} implies

\begin{equation} \label{sc2sc}
\begin{split}
\tfrac{1}{4} \check T_a{}^a \!-\!  \Lambda &=   \tfrac32 t^{2 \left(\alpha -1  \right)} \big[
\!-\! \alpha \left( \alpha \!+\!1 \right) 
 + 2 t \alpha \left( \alpha \!-\!1 \right) 
H^{\rm ext}
\\&
  -\tfrac{  t^2}{12} \left( \alpha -1 \right) \left( \alpha -2 \right) R^{g_{\bm \gamma}}
 +
 {\mathcal O}(t^4)\big]\, .
\end{split}
\end{equation} 
In the above $H^{\rm ext}:=-I_{\bm\sigma}\cdot I_{\bm \gamma}$ canonically  extends  the mean curvature of $\Sigma \hookrightarrow (M,g_{\bm \gamma})$.

As discussed earlier, solving Einstein's equations can be broken into two steps, (i) solve for a causal structure and (ii) determine which metric in the corresponding conformal class is physical. Hence we pose the question:
Given only  the trace of stress and causal structure for a spacetime with isotropic singularity, can we recover the
physical metric~$\check g\hh$? 
Remarkably there exists a ``solution generating algebra'' that 
addresses
 this question:
The  operator $I$ acting on~$\bm \sigma$ (of Section~\ref{IS})  is both conformally invariant and second order. It is an example of a more general  {\it Thomas $D$-operator} mapping tractors to tractors~\cite{BEG,Go}.
Given the data of a weight $w'\neq 0,-1$ density $\bm \rchi$, this yields 
a conformally invariant ``d'Alembert--Robin'' operator~\cite{f8}  \begin{multline}\label{dAR}
{\bm L}_{\bm \rchi} :=- w' \bm{\rchi } \left( \square +w J  \right) \\+ 2 \left(w+1 \right) \nabla_a \bm{\rchi} \nabla^a - \tfrac{w(w+1)}{w'+1} \left(\square \bm{\rchi} + w' J \bm{\rchi} \right)\, ,
\end{multline}
mapping weight $w$ densities 
to weight $w+w'-2$ densities.
When $g_{\bm \rchi} := {\bm \rchi}^{-\frac{2}{w'}}\bg$ is a metric,  this gives a d'Alembert operator
$
 \square^{g_{\bm \rchi}} +\tfrac{ w ( w+w'+2 )}{6(w'+1)}    R^{g_{\bm \rchi}} 
$.
Specializing $\bm \rchi$ to the singular Yamabe defining density $\bm \sigma$, the operator~${\bm L}_{\bm \sigma}$ yields a conformally invariant, 
Robin-type, boundary operator~\cite{Cherrier} 
$$\delta_{\rm R}\stackrel\Sigma{:=}\nabla_{\hat n} - w H\, .$$
The crucial point now is that, calling $
{\mathcal S}_{\bm \rchi} := 
{\bm L}_{\bm \rchi } {\bm \rchi} 
$,  there
 is an~$\mathfrak{ sl}(2)=\langle x,[x,y],y\rangle$ algebra generated by
 $$(x,y):=\big(\bm \rchi, -\tfrac{2(w'+1)}{w' {\mathcal S}_{\bm \rchi}}{\bm L}_{\bm \rchi}\big)\, .$$
Since  Eq.~\eqref{traceyourstress}  
can be rewritten as
$
 {\bm L}_{\bm \tau^{\alpha} } {\bm \tau^{\alpha}} =\tfrac{1}{3}
\check T_a{}^a - \tfrac43 \Lambda
$,  its  formal asymptotics 
can be determined iteratively using the solution generating~$\mathfrak{sl}(2)$ algebra, {\it cf.}~\cite{asym}.

\subsection{Conformal Fundamental Forms and Stress}

We now analyze the trace-free part of the matter coupled Einstein system in  terms of conformal embedding geometry. Eq.~\eqref{freeyourstress} implies that we must study the tractor gradient of Eq.~\eqref{formula}
%
%
relating the various scale tractors. 
 Acting with $q^*$ (see Eq.~\eqref{sctr}) and multiplying by $t^{-\alpha}$  gives \begin{multline}\label{eye2eye}
t^{-\alpha}q^*\bm\nabla I_{\bm \tau^{\alpha}}\! =
\tfrac{\alpha(\alpha-1)}{t^2} {\bm\gamma dt\circledcirc dt}
\\
+\tfrac{ \alpha }t q^*\bm\nabla 
I_{\bm \sigma}
+(1-\alpha) q^*\bm\nabla I_{\bm \gamma}
\, .
\end{multline}
Multiplying by an overall factor $2/\bm \gamma$, 
each (trace-free) term above has a physical interpretation:
The left hand side is
 the physical stress~\scalebox{.85}{$\mathring{\check T}$}. The first summand  is proportional to the stress of a perfect fluid with covelocity~$dt$. 
The second is $\alpha$ times the stress of the singular Yamabe metric. It captures the embedding data. The last  is $1-\alpha$ times the stress of the Big Bang metric~$g_{\bm \gamma}$. Hence we learn the asymptotics of trace-free stress~\scalebox{.85}{$\mathring{\check T}$} for all spacetimes with an isotropic singularity:
\begin{equation}\label{stressasym}
\mathring{\check T} = \frac{\alpha(\alpha-1)\mathring T_{\scriptscriptstyle\rm fluid}}{t^2} 
+ 
\frac{\alpha \mathring { T}_{\rm \scriptscriptstyle Bach}}t -
 (\alpha-1)\,  \mathring{T}_{\rm \scriptscriptstyle Big\hh Bang}\, ,
\end{equation}
where 
$
\mathring  T_{\scriptscriptstyle \rm fluid}:= 2d t\circledcirc d t
$.
Note that Eq.~\eqref{bach} implies
\begin{equation}\label{PnablaI}
 {\mathscr P} q^* \bm \nabla I_{\bm \sigma}
 =\bm \sigma B \stackrel\Sigma=0\, , 
 \end{equation}
so the (transverse order 2) partially massless operator acting on $\frac{\bm \gamma}2 \mathring T_{\rm Bach}$ returns~$\bm \sigma B$.

As advertized, Eq.~\eqref{stressasym} characterizes allowed stress at an isotropic singularity.  As we next show, the coefficients of terms that diverge as $t\to 0$ 
are local invariants of the boundary. 

\medskip
We want to study the first four orders of the early time  ($t\sim0$) asymptotics of  physical stress.
Both the fluid and Big Bang terms in Eq.~\eqref{stressasym} are completely determined
to this order so we  focus on the Bach term.
Conformally invariant transverse jets of~$q^*\bm \nabla I_{\bm \sigma}$ generate the  second and  third but not fourth conformal fundamental forms (see Eq.~\eqref{PnablaI}).
There is  a notion of a fifth fundamental form, {\it viz}  the projected Bach tensor $B^\top|_\Sigma$.
However the Bach-to-stress Eq.~\eqref{B2stress} determines the   conformal structure~$\bg$ given initial data 
of the first through fourth fundamental forms, so we focus on  these.

\medskip

First note  the second fundamental form here obeys
$$
\mathring K=q^* \bm \nabla I_{\bm \sigma}|_\Sigma
=\tfrac{\bm \gamma}2\, \mathring {T}_{\rm \scriptscriptstyle Bach}\big|_\Sigma
\, .
$$
To study the next order term, we use 
the tractor analog of the d'Alembert--Robin operator ${\bm L}_{\bm \sigma}$ of Eq.~\eqref{dAR} to make a transverse order 1 operator~\cite{CFF}, again called~$\delta_{\rm R}$, 
\begin{multline*}
\circledcirc T^*M[w]\ni
\mathring x_{ab}\stackrel{\delta_{\rm R}}\longmapsto
\big[
(\nabla_{\hat n} + (2-w)H)\mathring x_{ab}
\\\qquad\qquad
 +\tfrac2{w-3}\bar \nabla_{(a}\mathring x_{\hat n b)}^\top\big]^{\top,\circ}
\in \circledcirc T^*\Sigma[w-1]
\, .
\end{multline*}
The trace-free Fialkow tensor is then
$$
\mathring F = \delta_{\rm R} q^* \bm \nabla I_{\bm \sigma}
=
\delta_{\rm R}\big (\tfrac{\bm \gamma}2\, \mathring {T}_{\rm \scriptscriptstyle Bach}\big)
\, .
$$

Because we cannot extract $\mathring L$ 
 from a conformally invariant second normal derivative of $q^* \bm \nabla I_{\bm \sigma}$ to relate  the fourth fundamental form to stress, we instead consider one normal derivative of
 Big Bang stress $\mathring T_{\rm\scriptscriptstyle Big\hh  Bang}$. For this we employ the identity~\cite{f9}  
 $$
\delta_{\rm R}\hh q^*{\bm \nabla I_{\bm \gamma}}=
\bm \gamma \mathring L + \delta^{(2)} \bm \gamma\, .
$$
This yields the last line of Figure~\ref{fig} 
 summarizing the relations between geometry and stress.

\begin{figure}

\begin{tabular}{c|c}
Geometry & Stress\\\hline\hline\\[-3mm]
$
\hat n\circledcirc \hat n
$
& $\tfrac{{\bm \gamma}^2}2\mathring T_{\rm fluid}\big|_\Sigma$ \\[2mm]
$\mathring K$& $\tfrac{\bm \gamma}{2} \mathring{ T}_{\rm \scriptscriptstyle Bach}\big|_\Sigma
$
 \\[2mm]
$\mathring F$&
 $\delta_{\rm R}\big (\tfrac{\bm \gamma}2\, \mathring { T}_{\rm \scriptscriptstyle Bach}\big)$
 \\[2mm]
 $\mathring L$&
 $\bm \gamma^{-1}\delta_{\rm R}\big (\tfrac{\bm \gamma}2\, \mathring { T}_{\rm  
 \scriptscriptstyle Big\hh  Bang}\big)-\bm \gamma^{-1}{\sf \delta^{(2)}} \bm\gamma $
\end{tabular}
\caption{Conformal fundamental forms related to stress.}  \label{fig}
\end{figure}

\section{Example: Poincar\'e--Einstein 
Conformal Cyclic Cosmology}

Models where the present universe is seeded by pre-Big Bang data~\cite{ccc1,ccc2,ccc3,ccc4}, dovetail with the above results. One approach~\cite{cccn} employs
an asymptotically dS pre-Big Bang  
metric  $\hat g$ 
 and a physical metric $\check g$ with 
isotropic singularity:
$$
\hat g = \frac{-d\hat t^{\hh 2}+ \hat h(\hat t\hh )}{\hat t^2}\, ,\quad
\check g = \check t^{-2\alpha} \big (\!-d\check t^{\hh 2} + \check h(\check t\hh )\big)\, .
$$
The conformal infinity/initial singularity hypersurface $\Sigma$
is at  $\hat t=0=\check t$.
 The pre-Big Bang spatial metric $\hat h$ is  defined by a Fefferman--Graham-type expansion~\cite{FG} about the conformal infinity of $\hat g$ obtained by solving Einstein's equations with non-vanishing stress for suitable late time $\hat t\to 0_-$ matter content. 
 Conformal fundamental forms
are covariant analogs of Fefferman--Graham expansion coefficients~\cite{CFF} and are  determined by $\hat h(\hat t\hh )$.
 They can be matched~\cite{GoKo} to  those of the Big Bang model and thus its stress. Schematically,
$$
{\mathring{\hat { \hh T}}} \hh\mapsto \mbox{conformal fundamental forms}
\mapsto
\mathring{\check T}\, .
$$
 Just as for stellar models where interior and exterior solutions are matched using fundamental forms~\cite{OS,Darmois,Israel}, ``cyclic cosmological  matching'' of conformal structures is via conformal fundamental forms.  

\bigskip

\begin{acknowledgements}
 We thank Pawe\l\ Nurowski useful discussions.
A.R.G. and A.W. acknowledge support from the Royal Society of New Zealand via Marsden Grant 19-UOA-008. J.K.  acknowledges funding received from the Norwegian Financial Mechanism 2014-2021, project registration number UMO-2019/34/H/ST1/00636.
A.W.~was also supported by  Simons Foundation Collaboration Grant for Mathematicians ID 686131. J.K. and A.W. thank the University of Auckland for warm hospitality.

\end{acknowledgements}


\begin{thebibliography}{10}
\bibitem{f2}
We work in $d=4$ dimensions but all our considerations apply to dimension $d$ spacetimes $M$.
\bibitem{f4} 
A function $\tau$ is said to be defining for a hypersurface $\Sigma$ if $0\leq \tau\in C^\infty M$ and $\tau|_\Sigma = 0\neq d\tau|_{ p\in\Sigma}$ for all $p\in \Sigma$.
\bibitem{f5}
Transverse order counts normal derivatives. Also, the first conformal fundamental form is the boundary conformal structure.
\bibitem{f6}
The symbol $\top$ denotes projection along directions tangent to $\Sigma$. Also note there is no need to introduce new indices for hypersurface tensors if one identifies $T\Sigma$ with its image in $TM|_\Sigma$.
\bibitem{f7}
Remarkably, there is another canonical  metric 
$\big( |\nabla \bm \tau|_{\bm g}^2 -\tfrac{\bm \tau}{\alpha+1} (  \square \bm \tau +\tfrac1\alpha J^{\bm g}\bm \tau)\big)^{\frac\alpha{\alpha-1}}\, \bg$
defined to all orders in a collar neighborhood of spacelike~$\Sigma$, which also induces the metric $g_\Sigma$.  However, it is not directly related to conformal fundamental forms.
\bibitem{f9}
 Here $\delta^{(2)}:= -\mathring K \delta_{\rm R}^{(2)} - \mathring F \delta_{\rm R} + \bar \delta^{(2)}$ is a conformally invariant, transverse order 2 operator  on weight 1 densities. The operator $\delta_{\rm R}^{(2)}\stackrel\Sigma=\hat n^a \hat n^b \nabla_a\nabla_b + \cdots$
  is the second transverse order generalization  of the Robin operator of~\cite{LarryRod} (see Lemma 3.10 of~\cite{CFF} for an explicit formula; our conventions differ by a factor $3$). The boundary operator~$\bar \delta^{(2)}$ has transverse order 0  and is conformally invariant.
\end{thebibliography}
\end{document}